# Design and fabrication of dispersion controlled highly nonlinear fibers for far-tuned four-wave mixing frequency conversion

**SIDI-ELY AHMEDOU,** [1,*] **ROMAIN DAULIAT,** [1] **ALEXANDRE PARRIAUX,** [2] **ALIX MALFONDET,** [2] **GUY MILLOT,** [2,3] **LAURENT LABONTE,** [4] **SÉBASTIEN TANZILLI,** [4] **ROMAIN DALIDET,** [4] **JEAN-CHRISTOPHE DELAGNES,** [5] **PHILIPPE ROY,** [1] AND **RAPHAEL JAMIER**[1]

[1]*Université de Limoges, CNRS, XLIM, UMR 7252, F-87000 Limoges, France*
[2]*ICB, Université de Bourgogne Franche-Comté, CNRS, UMR 6303, F-21078 Dijon, France*
[3]*Institut Universitaire de France (IUF), 1 Rue Descartes, Paris, France*
[4]*Univ. Côte d'Azur, CNRS, Institut de Physique de Nice, F-06108 Nice Cedex 2, France*
[5]*CELIA, Centre Lasers Intenses et Applications, Université de Bordeaux-CNRS-CEA, UMR 5107, F-33405 Talence Cedex, France*
*\*sidi-ely.ahmedou@xlim.fr*

**Abstract:** We report on the conception, fabrication and characterization of a new concept of optical fiber enabling a precise control of the ratio between the 2$^{nd}$ and 4$^{th}$-order of chromatic dispersion (respectively $\beta_2$ and $\beta_4$) at 1.55 µm which is at the heart of the Four-Wave-Mixing (FWM) generation. For conventional highly nonlinear fiber the sensitivity of this ratio to fiber geometry fluctuations is very critical, making the fabrication process challenging. The new design fiber reconciles the accurate control of chromatic dispersion properties and fabrication by standard stack and draw method, allowing a robust and reliable method against detrimental fluctuations parameters during the fabrication process. Experimental frequency conversion with FWM in the new design fiber is demonstrated.

## 1. Introduction

In the last decades, the development of laser sources operating in the near-infrared (NIR) has been highly studied, which is mainly due to telecommunication applications and the use of silica fibers for signal transmission. In particular mature technological fiber lasers have been developed mainly in the spectral ranges corresponding to the emission spectra of rare-earth ions used for doping the fiber core. However, for other kind of applications such as spectroscopy, the NIR is not an ideal spectral region since most of molecules do not exhibit strong absorption features. This can be bypassed by operating in a wavelength domain above 2µm, and especially in the mid-infrared (MIR). Indeed, the MIR range corresponds to spectroscopic "fingerprints" of some important molecules such as $CO_2$, $CH_4$, CO or $NH_3$ which can be used for several applications such as medicine and environmental monitoring [1-4]. The availability of laser sources in this targeted spectral range is however limited. Thus, the development of coherent light sources operating in the MIR is attractive, and several possibilities have been investigated such as the use of quantum cascade lasers [5].

Frequency conversion techniques in nonlinear crystals [6] to reach higher wavelengths also present a strong interest since it keeps the features and qualities of the pump laser operating in the NIR. In the case of fibered lasers, the all-fibered feature can also be kept by using nonlinear phenomena inside a highly nonlinear fiber (HNLF), allowing long interacting lengths and a compact experimental setup. In particular, the wavelength conversion technique based on four wave mixing (FWM) in HNLFs can be used for frequency conversion with detuning of several tens of THz, and recent demonstrations even showed the possibility to reach the 2.6µm region [7,8]. However, the efficiency of this nonlinear interaction is based on a very accurate control and therefore knowledge of the chromatic dispersion and its higher orders derivative, notably the 2nd, 3rd and 4th derivative orders of the propagation constant β ($β_2$, $β_3$ and $β_4$ respectively) [9]. A precise control of these parameters at pump wavelength is then essential to ensure maximum conversion, especially when high detunings are targeted [10]. In fact, new frequencies can be efficiently generated in HNLFs by FWM if three conditions are simultaneously satisfied: a) the fiber is pumped close to the zero-dispersion wavelength (ZDW) in the normal dispersion regime, when nonlinear effects are neglected, b) the value of the fourth-order derivative ($β_4$) is negative, and c) the ratio $β_2/β_4$ is well controlled [7]. However, in this case, the phase-matching condition that guides the effectiveness of the FWM generation is sensitive to inevitable fiber geometry fluctuations (transversal and/or longitudinal) occurring during the drawing process of the preform giving birth to the fiber. While this sensitivity can appear like an advantage for tailoring dispersion curve around the pump wavelength, it can also be a serious drawback to precisely engineer the associated derivative $β_i$ of the propagation constant, as a slight deviation of theses parameters can lead to undesired effects [10].

There have been several reports on the control of chromatic dispersion using special fibers such as HNLFs, air-silica photonic crystal fibers (HNL-PCF) and highly nonlinear dispersion shifted fibers (HNL-DSF) [11-17]. Thanks to their opto-geometrical parameters, these fibers have been of great interest in tailoring waveguide dispersion and, through it, the chromatic dispersion properties. However, several inherent difficulties appear when tailoring associative derivative of the propagation constant in the spectral region beyond 1.55 µm. For instance, previous works have shown that in this wavelength range, the chromatic dispersion curve of PCF exhibits a complex shape, which complicates the control of the evolution of the $β_i$ parameters with respect to the fiber parameters [12,13]. A slight geometry fluctuation, as for instance the control of the air holes size during the fabrication process, will lead to abrupt dispersion properties change [11]. On the other hand, from the viewpoints of fabrication process, HNLF and HNL-DSF have been shown as a good alternative for both tailoring chromatic dispersion and reducing dispersion fluctuations owing to an easily controllable all-solid geometry [14,16]. While these fibers have been recognized as a powerful medium for tailoring chromatic dispersion, they introduce the fundamental limit of the high dispersion sensitivity in the vicinity of ZDW [18]. A slight transverse or longitudinal variation of the fiber geometry can drastically shift the targeted dispersion properties to an undesired wavelength. As an example, it has been shown that a deviation of 10 nm in core radius of conventional HNLF exhibits 0.35 ps/nm/km dispersion shifts from the targeted dispersion properties at a defined wavelength [19,20]. Unfortunately, state-of-the-art fabrication processes indicate that such a precision is very challenging to meet.

In this work, we propose an optimized optical fiber that allows a precise control of the two associated derivative $β_2$ and $β_4$ of the propagation constant and a precise control of their ratio $β_2/β_4$ at a given pump wavelength. Our motivation is the wavelength conversion by FWM of a frequency-agile all-fiber dual-comb [3] from the C- and L-bands (1.53-1.58 µm) to the 2 µm waveband with a seed wave tunable in the O-band (1.29 - 1.31 µm). Such a conversion is only possible with a high efficiency if an accurate control of the evolution of the dispersion parameters is met. For this purpose, a new concept of HNLF is proposed. Note that carbon dioxide absorption spectra in the 2 µm region show intense spectral lines of both $^{12}CO_2$ and $^{13}CO_2$ which could be used to measure the $^{13}C/^{12}C$ isotopic ratio.

## 2. Fiber Design

The FWM is a nonlinear process occurring in nonlinear medium such as an optical fiber [21,22]. Its efficiency is governed by the following phase–matching condition:

$$-4 \gamma P_p \leq \Delta\beta \leq 0 \quad (1)$$

Where $\gamma$ is the effective nonlinear coefficient, $P_p$ is the pump power of the light radiation travelling through the fiber core at the working wavelength and $\Delta\beta$ denotes the phase mismatch defined by:

$$\Delta\beta = 2\beta_2(\omega p - \omega)^2 + \beta_4(\omega p - \omega)^4/16 \quad (2)$$

As shown by Eq. (2) the phase-matching condition can be achieved by an appropriate design of the higher order dispersion $\beta_2$, $\beta_4$ ($\beta_2 > 0$ and $\beta_4 < 0$). And in this case the optimum frequency $f_{opt}$ for which the gain is maximal is given by:

$$f_{opt} = \frac{1}{2\pi}\sqrt{\frac{-2}{\beta_4}\sqrt{9\beta_2^2 - 6\beta_4\gamma P} - 6\frac{\beta_2}{\beta_4}} \cong \frac{1}{2\pi}\sqrt{\frac{-12\beta_2}{\beta_4}} \quad (3)$$

Moreover, the gain bandwidth is:

$$\Delta f \cong \frac{\gamma P_p}{2\pi\beta_2}\sqrt{\frac{-\beta_4}{3\beta_2}} \quad (4)$$

From Eq. (3) and Eq. (4), one can observe that the optimum frequency and the gain bandwidth are strongly sensitive to the ratio $\beta_2/\beta_4$. Thus, a precise control of this ratio, notably the control of $\beta_4$ will be crucial for the FWM efficiency. The dispersion properties which will guide the optical fiber design are:
- ZDW ≈ 1.56 µm (greater than the pump wavelength)
- -5800 ps$^{-2}$ < $\beta_2/\beta_4$ < -3650 ps$^{-2}$, delimited by tunability of the pump: 1.54 µm - 1.575 µm. The ideal target value is $\beta_2/\beta_4$ ~ -4900 ps$^{-2}$ for conversion shifts below 40 THz (1.56 µm towards 2 µm)
- $\beta_3 \ll 1$ to limit the undesired walk-off effect.

Most of time, dispersion-controlled fibers used for facilitating nonlinear effects like FWM in the 1.55 µm range are based on reduced fiber core size and increased numerical aperture in comparison with conventional fibers. In order to counteract the anomalous dispersion of silica in this spectral band, fiber designs often exhibit a W-type profile highlighting drastic variations of the mode field diameter due to the wavelength dependance of the mode field distribution guided in the fiber core. Thus, to fix the idea, one considers there an all-solid fiber as depicted in Fig. 1. The fiber core, exhibiting a graded index profile (the maximum refractive index value is noted $n_0$ at the center and the index difference is noted $\Delta n^+$ in comparison with pure silica), is surrounded by a thin low-index trench (refractive index noted $n_1$ and the index difference is noted $\Delta n^-$ in comparison with pure silica). The trench is surrounded by the optical cladding composed of pure silica (refractive index noted $n_g$). Thererafter, one considers that the high-index core is made of germanium-doped silica and the low-index trench is made of fluorine-doped silica.

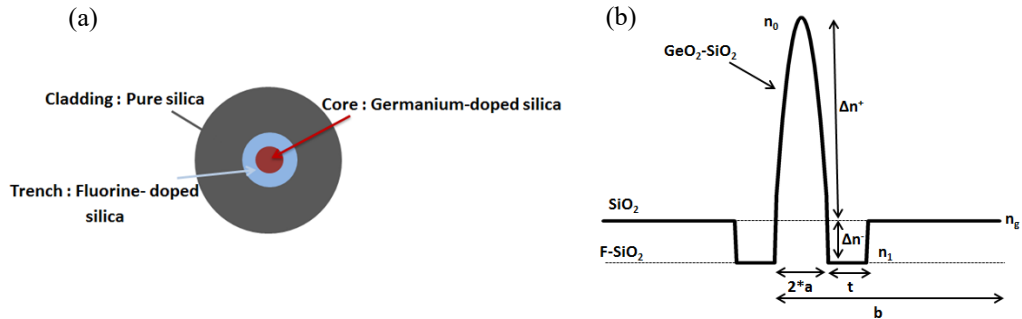

Fig. 1. (a) 2D cross-section of an all-solid W fiber. (b) Refractive index profile.

For short wavelengths, the electric field distribution of the fundamental mode is well-confined within the core thanks to the high refractive index contrast between the core and the trench. As wavelength increases, the electric field distribution will progressively extend outside the core and expand in the cladding area. As the mode intensity distribution steadily spreads out from the core, the resulting negative dispersion will thwart the anomalous dispersion of the silica material. At the vicinity of the ZDW, the waveguide dispersion of the fundamental mode has an inverted slope with respect to material dispersion. However, the ZDW and dispersion slope in this region become highly sensitive to geometrical fluctuations of the fiber design, which are unavoidable during the fabrication process [23]. While this sensitivity is required for tailoring the desired dispersion properties, it should be precisely controlled in order to ensure the efficiency of the FWM generation.

The chromatic dispersion is defined as the sum of the material and waveguide dispersions:
- Material dispersion ($D_M$): it results from the wavelength dependency of the refractive index of the glasses used for the fiber design. The fused silica dispersion is defined by Sellmeier's coefficients (the latter can be found in several databases such as [24]). The dispersion of Ge-doped silica and F-doped silica stand for Sellmeier's coefficients for $GeO_2$ glass (doping level fixed to 20 mole %) and coefficients of F-doped silica (doping level fixed to 1 mole %) [24].
- Waveguide dispersion ($D_W$): it arises from waveguide effect; it can be adjusted by optimizing the refractive index profile of the fiber [20].

It is worth reminding that the waveguide dispersion strongly depends on the core radius. Thereafter, as the outcome of a preliminary simulation campaign the core radius is fixed to a = 2.7 µm in order to obtain a ZDW near of 1.56 µm. The thickness of the trench can then be adjusted in order to optimize the position of ZDW and the dispersion slope close to the ZDW. Here we do not consider the influence of the refractive index contrast of the trench on the chromatic dispersion of the fiber, as we verified through simulation that the impact of varying the material composition by 1% is less significant than 1% variation in trench thickness. Typically, 1% in material composition will shift the ratio $β_2/β_4$ by a factor of 2000 and the ZDW position by 4 nm while 1% deviation in the trench thickness shifts the ratio $β_2/β_4$ by a factor of 8000 and the ZDW position by 16 nm. The chromatic dispersion is calculated using a commercial software based on the finite element method (COMSOL Multiphysics) giving access to the frequency dependent propagation constant ($β=2πn_{eff}/λ$). The chromatic dispersion (D) and its higher order derivative are given by the following relations:

$$D = \frac{-λ}{c}\frac{\partial^2 n_{eff}}{\partial^2 λ} \quad β_2 = \frac{-λ^2}{2πc}D \quad β_3 = \frac{-λ^2}{2πc}\frac{\partial β_2}{\partial λ} \quad β_4 = \frac{-λ^2}{2πc}\frac{\partial β_3}{\partial λ}$$

(5)

Where λ is the wavelength, c is the speed of light in vacuum and $n_{eff}$ is the effective index of the fundamental guided mode. Figure 2 displays the variation of the ratio $β_2/β_4$ and the ZDW with respect to the thickness (noted "t") of the low-n fluorine-doped trench, when $Δn^+$ is around 2.02 % and $Δn^-$ of 0.32%.

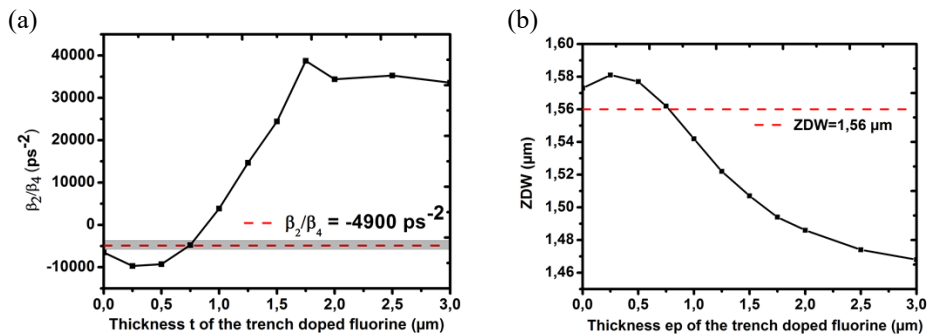

Fig. 2. Variation of (a) $β_2/β_4$ ratio calculated at λ = 1.55µm and (b) ZDW versus the thickness "t" of the low-n fluorine-doped trench surrounding the fiber core (as depicted in Fig.1). Here the core radius is fixed to 2.7 µm. The targeted values are highlighted by dashed red lines in both cases.

As it can be seen the ratio $\beta_2/\beta_4$ and the ZDW position are highly sensitive to the variation of the trench thickness. A slight variation of the order of 10 nm shifts drastically the $\beta_2/\beta_4$ ratio and the ZDW to a lesser extent from the targeted values, thus making their control during the fabrication process of the fiber very challenging, even impossible. To address this issue, the concept we propose relies on the discretization of the low-n trench, while retaining the same characteristics for the core (Fig.3).

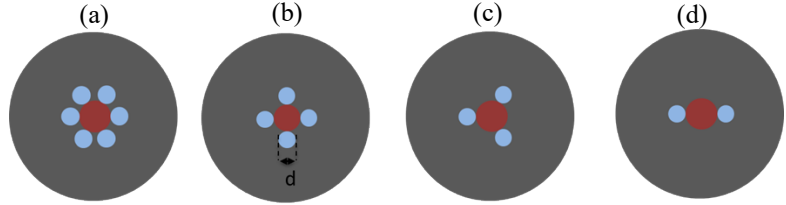

Fig. 3. New concept of fiber entitled D-HNLF in which the initial trench is discretized into either (a) six low-n inclusions (6I), (b) four low-n inclusions (4I),(c) three low-n inclusions (3I) or (d) two low-n inclusions (2I).

By doing so, one reduces the constraint imposed by the trench on the spatial electric field distribution of the guided mode and therefore gets a finer control over the mode spreading into the cladding as wavelength increases. Figure 4 shows the impact of discretizing the trench into inclusions (ranging from 2 to 6) on the slope of ratio $\beta_2/\beta_4$ and the ZDW position with respect to the diameter (noted "d") of the low-n inclusions. In order to facilitate the comparison with the fiber including the low-n trench, results previously depicted on Fig. 2 are added in Fig. 4 (dark curves).

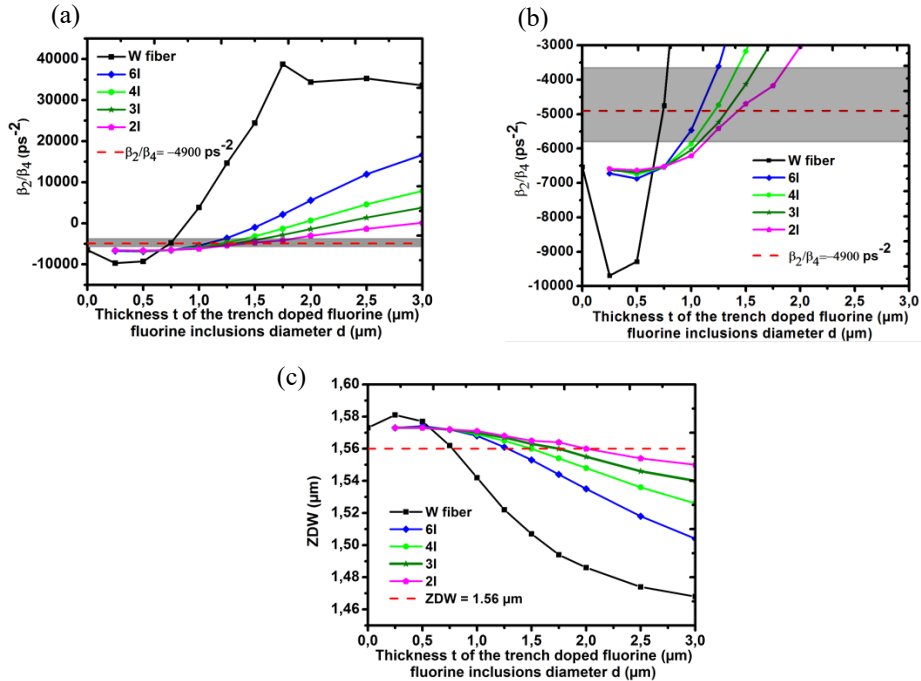

Fig. 4. Variation of (a) $\beta_2/\beta_4$ ratio calculated at $\lambda = 1.55\mu m$, (b) zoomed in view over the area of the interest defined by pump tunability and (c) ZDW versus the diameter "d" of the low-n inclusions surrounded the fiber core. Here the core radius is fixed to 2.7μm. Blue, light green, dark green and purple curves correspond respectively to the case where the discretized trench consists in 6, 4, 3 and 2 low-n inclusions. The targeted values are highlighted by dashed red lines in the two graphs. As a comparison, results obtained with the low-n trench and presented in Fig. 2 are added in the graphs (black curves).

It appears that, in comparison to the HNLF depicted in Fig.1 and based on a low-n trench, the use of a discretized trench composed of several low-n inclusions provides a smooth slope of the curves showing the evolution of the ratio $\beta_2/\beta_4$ and of the ZDW when the diameter "d" of the inclusions is modified. Moreover, it can be noticed that the slope of the different curves decreases as the number of low-n inclusions decreases (10483.4 ps$^{-2}$/µm for fiber with 6 inclusions to 2859.2 ps$^{-2}$/µm for fiber with 2 inclusions). Additionally, it has been noted that, for fibers with four, three and two fluorine inclusions, even until a variation of 400 nm in the fluorine inclusion diameter, the ratio $\beta_2/\beta_4$ and the ZDW position remains in the zone of the targeted parameters, thus revealing a good tolerance and flexibility compared to the trench thickness and the fiber with six fluorine inclusions. The benefit of a gentle slope is that it allows a precise tuning of the $\beta_2/\beta_4$ ratio and the ZDW position as well as a good flexibility during the fabrication process.

## 3. Fabrication and measurement

The four, three and two inclusions "D-HNLF" proposed in the previous section have been fabricated and characterized as they offer the best trade-off between a precise control of $\beta_2/\beta_4$ ratio and an improved tolerance to the manufacturing process fluctuations as well. These fibers have been fabricated by the well-known stack and draw process, using different rods: a germanium-doped rod exhibiting a parabolic refractive index profile for the core, pure silica ones for the optical cladding and fluorine-doped silica rods for the low-n discretized trench. To prevent the creation of parasitic bubbles during drawing, a thin layer (with a thickness of 0.83 µm) of pure silica surrounds the Ge-doped core. We therefore considered the presence of this silica layer in our theoretical model in order to take into account its impact on the $\beta_2/\beta_4$ ratio and the ZDW position. The presence of this silica layer, frontier in between the core and the low-n inclusions constituting the discretized trench, induces a slight variation of the values obtained for the $\beta_2/\beta_4$ ratio and the ZDW position. However, this slight modification can be easily compensated by reducing the low-n inclusion diameter. The following table summarizes the theoretical parameters used for the fabrication of the fibers.

Table 1. Fibers theoretical targeted parameters at 1.55 µm

| Parameters | Four inclusions | Three inclusions | Two inclusions |
|---|---|---|---|
| Core | 2.7 µm | 2.7 µm | 2.7 µm |
| d (µm) | 0.9 | 1 | 1.3 |
| ZDW (µm) | 1.5668 | 1.567 | 1.566 |
| $\beta_2$ (ps$^2$/km) | 0.93 | 0.963 | 0.934 |
| $\beta_3$ (ps$^3$/km) | 0.0680 | 0.07 | 0.0703 |
| $\beta_4$ (ps$^4$/km) | $-1.917 \times 10^{-4}$ | $-1.925 \times 10^{-4}$ | $-1.93 \times 10^{-4}$ |
| $\beta_2/\beta_4$ (ps$^{-2}$) | -4845.37 | -4999.43 | -4885.87 |
| MFA (µm$^2$) | 16.43 | 16.43 | 16.43 |

Although the stacking/drawing of solid canes fixes the proportion between the core and inclusions diameters, each fiber was drawn with controlled variations on the outer diameter in order to get a set of samples with small variations of dimensions and associated properties. The group delay ($GD_r$) has been measured over a wide spectral band (1.15 µm to 1.65 µm) using a Mach-Zehnder interferometer based on a white-light source [25]. By fitting the experimental data with a polynomial of the 4$^{th}$ order, the dispersion is thus deduced using the following equation.

$$D = \frac{\partial GD_r}{\partial \lambda} \tag{6}$$

In Fig. 4(a), the experimental data (black dots) are fitted (red line) using polynomial of order 4. For comparison, we also reported the theoretical dispersion curve (dark lines) of the two inclusions fiber on Fig. 4(b), the three inclusions fiber on Fig. 4(c) and the four inclusions fiber on Figure 4(d). Overall, the experimental curves plotted in Fig. 4(b, c, d) (red lines) are consistent with the predictions of the theoretical model particularly in the region of interest (1.5 µm to 1.6 µm). However, we notice an offset between the two curves in the range form 1.2 µm to 1.45 µm, which we can explain by the wide measurement interval in this region. Fibers dispersion properties of the fabricated fibers are summarized in Table 2. Thanks to the good agreement between theoretical and experimental curves, the focus will henceforth be on the frequency generation by FWM.

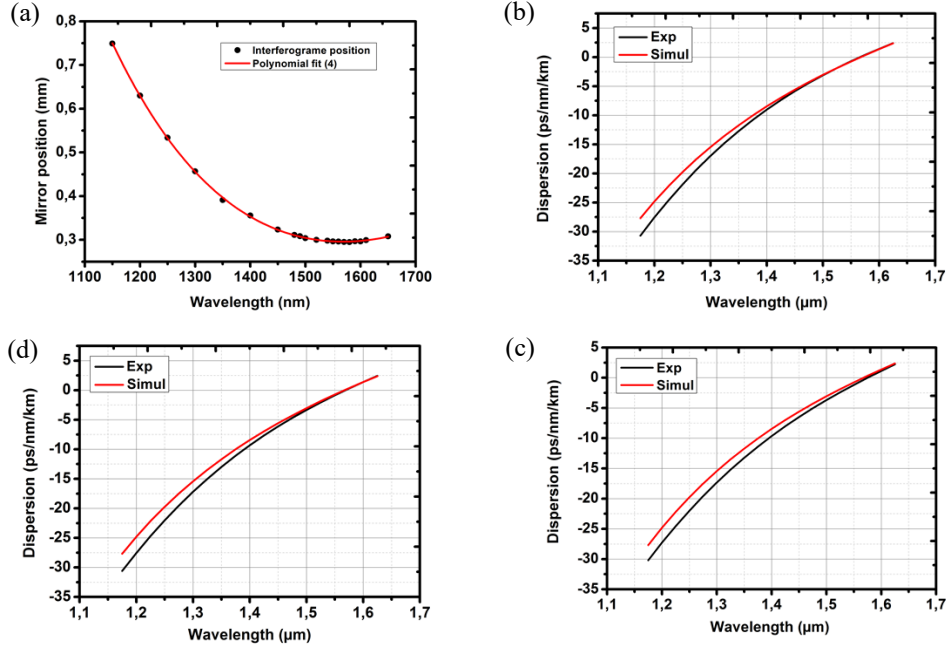

Fig. 5. (a) Experimental data fitting with polynomial of order 4. Chromatic dispersion curves of fiber with (b) two fluorine inclusions (c) three fluorine inclusions and (d) four fluorine inclusions

Table 2. Fibers measured dispersion parameters at 1.55 µm

| Parameters | Four inclusions | Three inclusions | Two inclusions |
|---|---|---|---|
| ZDW (µm) | 1.567 | 1.575 | 1.574 |
| $\beta_2$ (ps$^2$/km) | 0.84 | 0.948 | 0.979 |
| $\beta_3$ (ps$^3$/km) | 0.074 | 0.0746 | 0.0712 |
| $\beta_4$ (ps$^4$/km) | -1.73*10$^{-4}$ | -2.15*10$^{-4}$ | -2.14*10$^{-4}$ |
| $\beta_2/\beta_4$ (ps$^{-2}$) | -4855.49 | -4409.3 | -4574.76 |

An experimental demonstration of frequency generation with FWM using 60 meters of a three inclusions D-HNLF was also performed. We used as a pump a train of pulses with a repetition frequency of 100 MHz and an average power of 130 mW, whereas the seed is a 30 mW continuous laser. The wavelengths of the pump and the seed are respectively 1562 nm and 1300 nm, in order to generate a signal around 1950 nm, in the absorption window of $CO_2$. The results, depicted in Fig. 6, are compared to the ones obtained with the same length of a commercial HNLF. The latter has a nonlinear coefficient γ of 20 W$^{-1}$.km$^{-1}$, a dispersion coefficient D of -0.6 ps/nm/km, a slope $S$ of 0.026 ps/nm$^2$/km, and a ZDW of 1575 nm. The polarization has been optimized with both the D-HNLF and the HNLF, in order to make a comparison under similar conditions. Because the splicing losses are different from one fiber to another, the powers have also been adjusted, so that input powers of the pump and the seed are the same for both fibers. We can see that the signal generated with the D-HNLF (Fig. 6 (a)) is improved by 9 dB compared to the one obtained with the commercial HNLF (Fig. 6 (b)). We can also notice that the result with the HNLF exhibits Raman peaks, which are not present with the D-HNLF. However, these peaks are too weak to cause a significant pump depletion. These preliminary experimental results are very encouraging and validate the interest of our new fiber design.

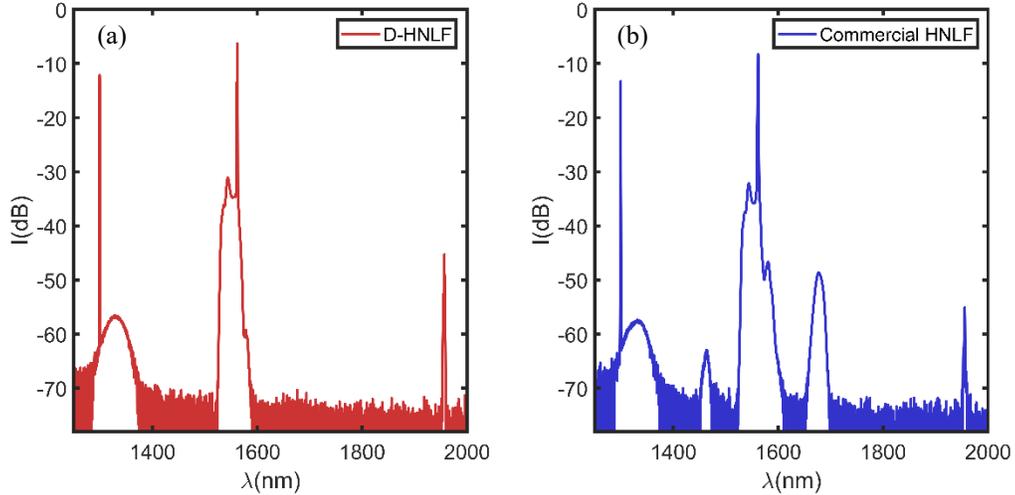

Fig. 6. Experimental spectrum recorded at the output of a three inclusion D-HNLF (a) and of a commercial HNLF (b).

## 4. Conclusion and perspectives

This work reported on the conception, the fabrication and the characterization of a new concept of optical fiber for precisely tailoring higher order dispersion. Indeed, $\beta_2$ and $\beta_4$ play a critical role for conversion based on MI and therefore, their values at the pump wavelength need to be accurately controlled. It has been demonstrated that in contrast to the standard HNLF, the new design allows to precisely control the $\beta_2/\beta_4$ ratio in the vicinity of the cut-off wavelength when the dispersion properties are highly sensitive to the geometry fluctuation. In a preliminary simulation investigation, we have shown that for HNLF a slight variation of the thickness of the trench surrounding the core can shift abruptly the ideal $\beta_2/\beta_4$ ratio required for an efficient conversion near mid-IR. Whereas, for the same order of magnitude variation as the trench thickness, the new D-HNLF fiber which relies on discretizing the trench provides a better control of dispersion properties at the targeted wavelength 1.55 μm. Consequently, it allows a better flexibility during the manufacturing process, and therefore to be able to draw a long length of fiber without compromising the desired dispersion properties.

Structures with four, three and two fluorine inclusions of the new D-HNLF fiber have been manufactured and measured. Experimental results have shown a good agreement with the calculations. The frequency conversion near-IR in these fibers has been successfully demonstrated.

**Funding.** The authors would like to thank the Agence Nationale de la Recherche (ANR) for the financial assistance it provided for the METROPOLIS/ANR-19-CE47-0008 project and the thesis of Sidi-Ely Ahmedou.